\documentclass[preprint,showpacs,prb]{revtex4}
\usepackage{stmaryrd}
\usepackage{amsmath}
\usepackage{amssymb}
\usepackage{graphicx}
\usepackage{dcolumn}
\usepackage{bm}

\begin{document}

\begin{titlepage}

\title{Gate-independent energy gap in non-covalently intercalated bilayer graphene on SiC(0001)}

\author{Yuanchang Li\footnote{liyc@nanoctr.cn}}
\address{National Center for Nanoscience and Technology, Beijing 100190, People¡¯s Republic of China}

\date{\today}

\begin{abstract}
Our first-principles calculations show that an energy gap around 0.12-0.25 eV can be engineered in epitaxial graphene on SiC(0001) through the non-covalent intercalation of transition- or alkali-metals, yet originated from the distinct mechanisms. The former is attributed to the combined effects of metal induced perpendicular electric field and interaction while the latter is solely attributed to the built-in electric field. A great advantage of this scheme is that the gap size is almost independent of the gate voltage up to 1 V/nm, thus reserving the electric means to tune the Fermi level of graphene when configured as field-effect transistors. Given the recent progress in experimental techniques on the intercalated graphene, our findings provide a practical way to incorporate the graphene with the current semiconductor industry.
\end{abstract}


\maketitle

\draft

\vspace{2mm}

\end{titlepage}
The lack of an energy gap represents a major bottleneck for the applications of graphene in post-silicon electronics and photonics although its many superior properties\cite{Novoselov}. In this regard, bilayer graphene holds more promise because a turnable gap has been realized under an external electric field which leads to the asymmetry of top and bottom layers (see Fig. 1)\cite{Zhang,MakPRL,Castro,Weitz,McCann}. Likewise, molecular doping to bilayer graphene also yields the charge distribution asymmetry between top and bottom layers, inducing an interlayer electric field and hence the energy gap\cite{YuWJNL,ZhangACSn}. However, in practice, a perpendicular electric field, i.e., the gate, is necessarily employed to tune the graphene Fermi level when graphene is configured as field-effect transistors. As such, the sophisticated dual-gate mode has to be invoked in order to achieve the independent control over the carrier concentration and the gap tuning\cite{Weitz}. Therefore, an energy gap opened intrinsically instead triggered by a gate voltage or chemical doping is highly desired from a technological point of view. On the other hand, a substrate with good thermal and mechanical stabilities but minimum degrading graphene electronic property is another prerequisite for its applications in electronics\cite{usprl}.

To this end, an ideal scheme is to develop a built-in electric field across the bilayer graphene through the underlying substrate engineering. Epitaxial graphene on SiC(0001)\cite{deHeer} provides an excellent platform not only because graphene with high quality can be directly synthesized therein but also because there exists an interfacial buffer layer between the true graphene and SiC which allows for the local property modulations with little damage to the atop graphene. For example, the quasi-free-standing graphene\cite{Riedl} and room-temperature ferromagnetism\cite{cpcprb,Giesbersprl} have been experimentally reported through the hydrogen intercalation. Intercalation of first-layer graphene will lead to the bilayer product due to the decoupling of buffer layer from the SiC substrate. A remarkable consequence of this processing is to leave an exotic element decorated SiC. Previous studies have shown that a sufficiently built-in electric field would be induced for the metal-decorated fullerenes\cite{usB80jpc,Caprl}. If this is also true for metal-decorated SiC, the bilayer graphene located above must be subjected to a vertical electric field which would result in the gap opening spontaneously. Nonetheless, the gap opening should be not so efficient for a covalent intercalation such as H or N elements.

\begin{figure}[tbp]
\includegraphics[width=0.75\textwidth]{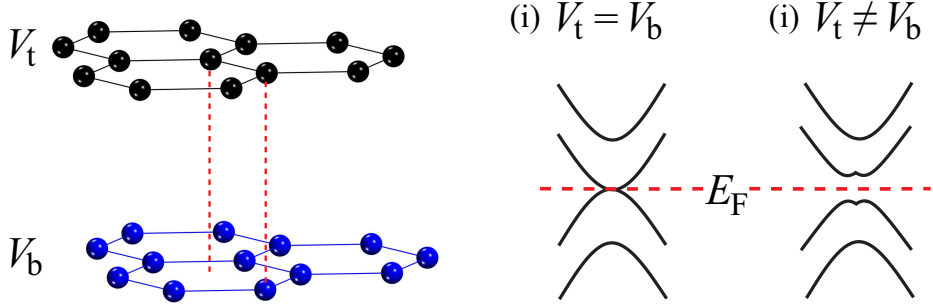}
\caption{\label{fig:fig1} (Color online) Schematics of the geometric and electronic structures of $AB$-stacked bilayer graphene. $V_t$ and $V_b$ denote the electric potentials of top and bottom graphene layers, respectively. When (i) $V_t$ = $V_b$, it shows a gapless band structure with the parabolic dispersion. When (ii) $V_t$ $\neq$ $V_b$, an energy gap is opened, showing a ``Mexican-hat" dispersion.}
\end{figure}

In this paper, we show that it is possible to generate a gap around 0.12-0.25 eV in epitaxial first-layer graphene on SiC(0001) via non-covalent intercalation such as transition- and alkali-metals by using first-principles calculations. The intercalated metal atoms break the interfacial Si-C bonds, consequently recovering the graphene feature of the carbon buffer layer, which leads to the configuration of bilayer graphene located on metal-decorated SiC. For alkali-metal(AM), it produces a perpendicular electric field and thus gives rise to an energy gap of 0.25 eV below the Fermi level. While for transition-metal (TM), apart from the electric-field gap, an additional gap $\sim$0.12 eV is found owing to the strong coupling between bottom graphene layer and the intercalated atoms which also brings about the symmetry-breaking of bilayer graphene. The $d^5$ TM (Mn, Tc and Re) systems are special cases where the Fermi level falls just in the bandgap. More importantly, the equivalent built-in electric field is so high, up to 4 V/nm, that the magnitude of gap is almost insensitive to the gate tuning ($\leq$ 1 V/nm), behaving like an intrinsic one. Put together, our system may simultaneously realize the gap opening, carrier tuning and compatible substrate in graphene, and therefore open up a way to incorporate graphene with the current semiconductor industry.

All calculations were performed within the framework of density functional theory (DFT) using the Vienna \emph{ab-initio} simulation package (VASP) \cite{vasp}. The local spin density approximation (LSDA) \cite{CA} was employed to describe the exchange and correlation effects of electrons. The electron-ion interaction was described by a projector augmented wave method \cite{PAW} with an energy cutoff of 400 eV. A gamma-centered grid of 12 $\times$ 12 $\times$ 1 was used to sample the Brillouin zone. The slab model is adopted with a $\sqrt3 \times \sqrt3R30^{\circ}$ supercell for 6$H$-SiC(0001) which can accommodate a 2 $\times$ 2 graphene cell and one intercalated atom. The SiC substrate contains six SiC bilayers with hydrogen passivation of the bottom surface. The lower half of SiC is fixed at its respective bulk position in the simulation while the other atoms are relaxed until the residual forces are less than 0.01 eV/\AA. For the hybrid density functional (HSE06) calculations, only three SiC bilayers are used as the substrate because of its heavy computational cost. Test calculations show that such a treatment yields an excellent description of the states around the bandgap.

\begin{figure*}[tbp]
\includegraphics[width=0.99\textwidth]{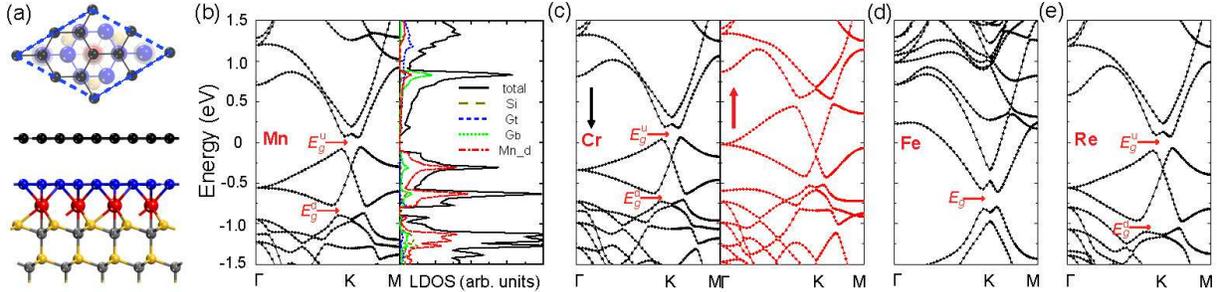}
\caption{\label{fig:fig2} (Color online) (a) Top and side view of the optimized configuration of Mn intercalated system. Yellow and red balls are Si and Mn, respectively, while others are C locating at different positions. Rhombus denotes the supercell. For simplicity, only the two topmost SiC bilayers are shown. (b)-(e) Typical electronic structures corresponding to different TM intercalation, (b) Mn, (c) Cr, (d) Fe, and (e) Re. Total density of states as well as the partial contributions from bottom (G$_b$) and top (G$_t$) graphene layer, intercalated Mn $d$-orbitals and surface Si atoms are shown in the right panel of (b). Fermi levels are set to zero. In (c), the black and red lines denote down- and up-spin states, respectively.}
\end{figure*}

We first investigate the intercalated systems with 3$d$ TMs from Sc to Ni. Figure 2(a) plots the typical geometric configuration by taking Mn as an example. It is found that in the ground state, the bilayer graphene always presents the AB-staking and the TMs lie at the interface between bilayer graphene and SiC substrate. For example, the AA-stacking configuration for Mn intercalated system is 66 meV less stable as compared with the AB-stacking one. We also calculate the energetics of the three possible positions for Mn, namely, adsorbed on, located between and underneath the bilayer graphene. It is revealed that the Mn underneath configuration is energetically more stable by 3.54 and 0.26 eV than the first two cases. This should be the reason why TMs prefer to penetrating through the graphene lattice to the interface under thermal annealing as experimentally observed\cite{Gao,Yagyu,Wang,Lima}. Mostly, the TMs are located exactly under the hexagon center of bottom graphene with the exception of Co and Ni in which the spontaneous symmetry-lowering occurs just like the situation of the single layer graphene\cite{usprl}. In contrast, with respect to SiC, the TMs always bind to three surface Si atom with a local $C_{3v}$ symmetry and thus fully saturate the Si dangling bond states.

Then let us explore their electronic properties, which are selectively shown in Figs. 2(b)-2(e) corresponding to the systems of TM = Mn, Cr, Fe, and Re. In fact, there appears three typical bands associated with different TMs, that is, (i) Sc and Ti systems have no gap everywhere (not present here), (ii) V, Cr and Mn systems exhibit two gaps ($E_g^u$ and $E_g^d$) around the Fermi level [see Figs. 2(b) and 2(c)], (iii) Fe, Co and Ni systems have only one gap below the Fermi level, meaning an $n$-type doping [see Fig. 2(d)]. Furthermore, the TMs in the same column of the periodic table give rather similar band characteristic as manifested by the comparison between Mn and Re systems [see Figs. 2(b) and 2(e)] which both belong to the $d^5$ TMs. This is also true for the $d^4$ and $d^6$ TMs. Such a phenomenon seems to be prevalent in all the TM intercalated systems\cite{jpcQAH}. In particular, all the $d^4$ TM systems are spin-polarized. For example, the Cr intercalated system has an almost integer magnetic moment of 0.93 $\mu_B$ as shown in Fig. 2(b). A recent work\cite{Seixas} showed that the ferromagnetism and ferroelasticity might coexist in materials with a Mexican-hat-like band, forming a stable multiferroic two-dimensional systems. Note that our systems are all characterized by the Mexican-hat-like band and even the inherent ferromagnetism in the Cr system. In addition, it is the transition metal $d_{x^2-y^2}$ and $d_{xy}$ states\cite{usprl, usjpcm}, having the same symmetry as for $d$-wave superconductivity in doped graphene, that characterize the electronic properties of the intercalated graphene, which may lead to the superconductivity in high temperature\cite{Schaffer}. These suggest the possibilities to realize the much sought-after multiferroic property in our system.

The Mn intercalated system is taken as an example to further explore the electronic property, whose local density of states (DOS) are plotted in the right panel of Fig. 2(b). It can be seen that the states around the Fermi level are mainly contributed by the Mn $d$-orbitals and the bottom graphene layer, implying their strong hybridization and hence endowing the system an excellent thermal and mechanical stability.

\begin{figure}[tbp]
\includegraphics[width=0.8\textwidth]{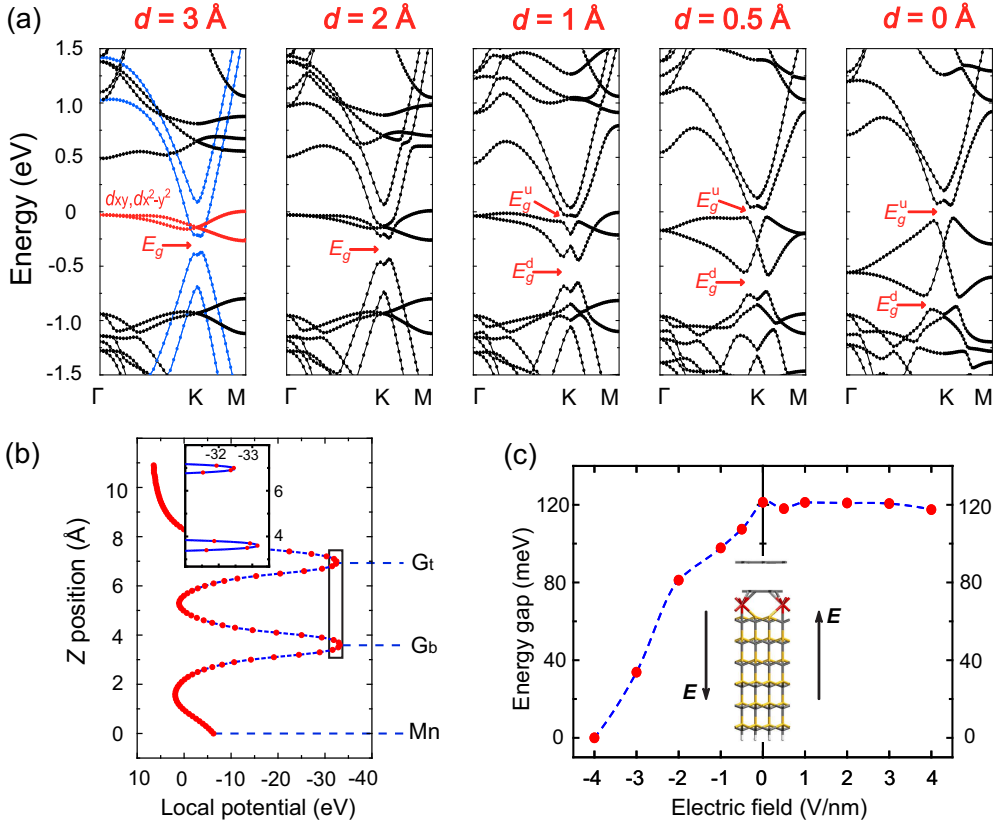}
\caption{\label{fig:fig3} (Color online) (a) Evolution of the band structure of Mn intercalated system as a function of the vertical separation $d$ between the bottom graphene and Mn. Zero is the Fermi level. Here $d$=0 corresponds to the ground state. At $d$=3 \AA\, the red and blue lines denote the states dominantly contributed by Mn ($d_{xy}$, $d_{x^2-y^2}$) doublet and bilayer graphene, respectively. (b) Local potential of Mn intercalated system at $d$=2 \AA\, where Mn, G$_{\rm b}$ and G$_{\rm t}$ denote the positions of intercalated Mn, bottom and top graphene layer. (c) Dependence of the $E_g^u$ on the applied electric field.}
\end{figure}

Next we turn to the focus of this paper, i.e., the gap opening. In order to unravel the underlying physics, we track the band structure evolution as the function of perpendicular separation ($d$) between bilayer graphene and Mn-decorated SiC. Figure 3(a) gives the results corresponding to $d$ = 3, 2, 1, 0.5 and 0 \AA\ (Note that $d$ = 0 corresponds to the ground state). At $d$ = 3 \AA, the vertical separation between Mn and bottom graphene is up to 4.5 \AA\ so that there must be no direct bonding between them. Surprisingly, an energy gap ($E_g$) from -0.38 eV to -0.22 eV has appeared already. The band character analysis reveals that the six bands around the Fermi level are dominantly contributed by the Mn $d_{xy}$ and $d_{x^2-y^2}$ orbitals (red lines), as well as the bilayer graphene (blue lines).

At $d$ = 2 \AA, the $E_g$ still exists but its size enlarges a little, to 0.20 eV. At the same time, the hybridization between bilayer graphene and Mn $d$-orbitals becomes more significant. When $d$ = 1 \AA, there emerges a new bandgap (denoted as $E_g^u$) and the original bandgap $E_g$ (denoted as $E_g^d$ hereafter) evolves to 0.21 eV. Further decreasing $d$, the $p$-$d$ hybridization is stronger and stronger and as a result, the Mn $d_{xy}$ and $d_{x^2-y^2}$ bands become more and more dispersive. However, the feature of $E_g^u$ and $E_g^d$ coexistence remains the same.

A natural question is the origin of $E_g^u$ and $E_g^d$. Above all, the gap opening in bilayer graphene must arise from the symmetry-breaking between the top and bottom graphene layer, as schematic illustration in Fig. 1. Because $E_g^u$ only emerges after a critical $d$ it should correspond to a short-ranged symmetry-breaking mechanism while $E_g^d$ is always there, thus corresponding to a long-ranged symmetry-breaking mechanism. Previous studies have shown that the metal-decorated fullerenes can give rise to an effective electric field along the radial direction due to the substantial charge redistribution upon metal decoration\cite{Caprl,usB80jpc}. Similarly, the TM-decorated SiC could also produce an electric field but instead along the vertical direction given the unique geometry in our systems. In order to illustrate this, we calculate the local potential of $d$ = 2 \AA\ case and the result is plotted in Fig. 3(b). It is found that the local potential does become distinctly different at the positions of top and bottom graphene layers, with the difference about $\Delta$ = 0.7 eV. Further taken into account the vertical separation of $h$ = 3.35 \AA\ between the bilayer graphene, we can roughly estimate the induced electric field ($E$) by $\Delta$ = $eEh$. It gives $E$ = 2.1 V/nm. As demonstrated previously, such an electric field would yield an energy gap around 0.25 eV\cite{MakPRL,guoapl}, consistent well with here obtained 0.20 eV. Note that such an electric field essentially arises from the metal decoration on SiC and thus exists irrelative with the separation $d$. Accordingly, it must be responsible for the $E_g^d$.

Another way to break the inversion symmetry in bilayer graphene is to introduce interaction\cite{Nandkishore,Freitag}, e.g., the bonding which works locally unlike the electric field. The fact that the $E_g^u$ only appears when $d <$ 1 \AA\ unambiguously indicate the key role of the hybridization between SiC modulated Mn $d$-orbitals and bilayer graphene in the $E_g^u$ formation. This is also evidenced by the synchronous evolution of $E_g^u$ and the broadening of $d_{xy}$ and $d_{x^2-y^2}$ states as shown in Fig. 3(a). Previous studies\cite{usprl,usprbRap} on Mn-decorated SiC told that its three $z$-related $d$-orbitals fully saturate the three Si dangling bond states, leaving the $d_{xy}$ and $d_{x^2-y^2}$ orbitals atomic-like in the vicinity of Fermi level which then heavily hybridize with C $p_z$ orbitals. As the $d$ decreases, the hybridization of Mn with bottom graphene becomes stronger and stronger so that the two originally narrow $d$-bands disperse more and more along with the repulsion of lower part of bilayer graphene states to deeper energy. Such a $p$-$d$ hybridization leads to the potential difference at the bottom and top graphene layers, thereby breaking the inversion symmetry and yielding a new gap $E_g^u$ at the Fermi level besides $E_g^d$.

It is noted that the $p$-$d$ hybridization and the induced electric field compete with each other. This is directly reflected by the change of $E_g^d$ as the decrease of $d$: the $E_g^d$ first increases from 0.16 eV, reaches a maximum of 0.21 eV at $d$ = 1 \AA\ and then decreases to 0.09 eV at $d$ = 0 \AA. In addition, the effect is TM species dependent in terms of their different number of $d$-electrons. For example, the $E_g^d$ and $E_g^u$ are at different energies in the Mn and Cr systems [see Figs. 2(b) and 2(c)] while they are at the similar energy in the Fe system where the two combines together into an overall gap [see Fig. 2(d)]. Such an interplay between $E_g^u$ and $E_g^d$ is also revealed by our calculations using hybrid density functionals on the Mn system. As expected, the $E_g^u$ is increased from 0.12 eV of LSDA to 0.18 eV of HSE06, however, the $E_g^d$ is surprisingly decreased from 0.09 eV to 0.07 eV.

In current semiconductor industry, the field-effect transistor based devices require a gate electric field to tune the system Fermi level within the sample. It is thus of great importance to explore the gating effect on the system gap because the applied electric field may affect the built-in one either positively or negatively. Figure 3(c) plots the dependence of $E_g^u$ [see Fig. 2(b)] on the applied electric field. It can be seen that under a positive gate (along the direction of SiC-Mn-bilayer), the gap varies just a little, keeping around 0.12 eV even if the field strength reaches extremely high, 4 V/nm. In contrast, under a negative gate, the gap diminishes with the increased field strength, and is thoroughly closed at -4 V/nm. Accordingly, we deduce that the built-in electric field by Mn-decorated SiC is equivalent to a positive gate of 4 V/nm. It is worth to note that the electrical breakdown field limit is only 0.4 V/nm and 1.0 V/nm, respectively for SiC\cite{Harris} and SiO$_2$\cite{Castro} substrates while the $E_g^u$ maintains larger than 0.1 eV for the gate ranging from -1 to 1 V/nm. This means the gap will be largely independent of the gate tuning, behaving like an intrinsic one which certainly allows the conventional CMOS technology to work. Significantly, a very recent work\cite{HaoYF} has shown that the 0.1 eV gap is enough to yield an on/off ratio greater than 10$^4$ in a graphene-based device. Therefore, here reported design scheme signifies the high potential of TM intercalated SiC systems for applications in future graphene electronics.

\begin{figure}[tbp]
\includegraphics[width=0.65\textwidth]{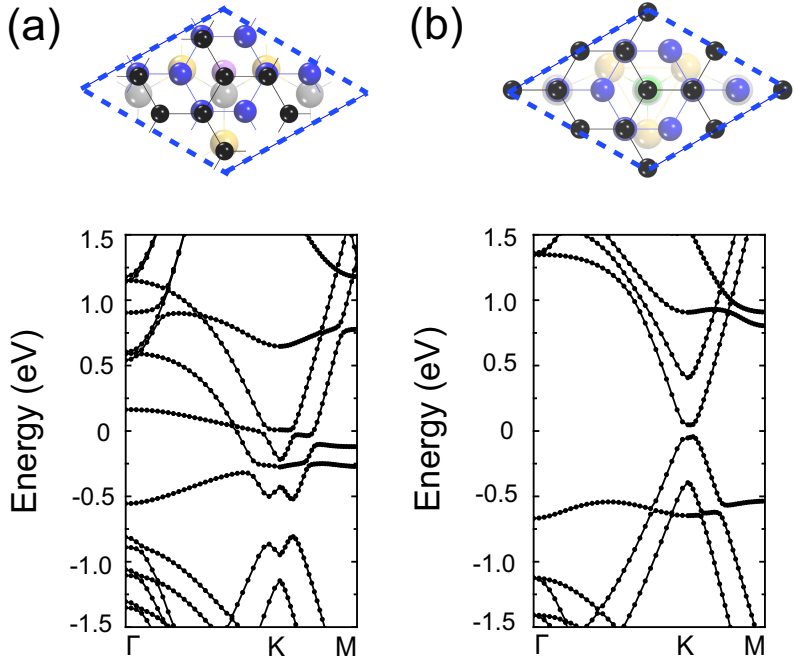}
\caption{\label{fig:fig4}(Color online) Geometric and electronic band structures of (a) Li, and (b) N intercalated bilayer graphene on SiC(0001). Yellow, pink, and green balls are Si, Li, and N, respectively, while others are C locating at different positions. Rhombus denotes the supercell. Fermi levels are set to zero.}
\end{figure}

Note that there also exists the gap $E_g^d$ which is totally originated from the induced perpendicular electric field by TM-decorated SiC. As known to us, the AMs\cite{usB80jpc} can generate such an electric field too and a similar gap would be anticipated. In experiment, the AMs including Li\cite{Virojanadara,usLijpc} and Na\cite{XiaXNa,SungNa} have also been successfully intercalated into the interface between graphene and SiC substrate. In Fig. 4(a), we show the geometric and electronic structure of Li intercalated system. Now the Li only directly binds to two surface Si atoms and does not bond to the carbon in graphene, essentially different from that of TM intercalation. Nevertheless, an energy gap appears as well between [-0.81, -0.56] eV although its indirect feature from $K$ to $\Gamma$. Here observed gap opening solely stems from the metal decoration induced electric field given the relatively large vertical separation (1.9 \AA) of Li apart from bilayer graphene and also their weak coupling\cite{usLijpc}. For comparison, we further explore the case of covalent intercalation using N element which is also experimentally realized\cite{WangZJ}. Figure 4(b) shows the optimized geometric configuration and the corresponding band structure. An energy gap $\sim$0.08 eV comes out across the Fermi level. The gap size is reduced significantly compared with 0.25 eV in Li intercalated system. This is not only due to the much larger vertical separation between N and bilayer graphene (2.9 \AA) but also because the tiny charge transfer between them.

Before closing, let us briefly discuss the experimental feasibility. Although the smaller $\sqrt3 \times \sqrt3R30^{\circ}$ reconstruction has also been observed on SiC(0001)\cite{Forbeaux,Rutter,Tromp}, the usually observed is the $6\sqrt3 \times 6\sqrt3R30^{\circ}$ reconstruction accommodating a $13 \times 13$ graphene cell. This leads to the displacement of intercalated TM atoms away from the graphene hexagonal center and therefore affects the coupling between them. However, we believe that the gap opening should always exist because of the inevitable symmetry-breaking in bilayer graphene, i.e., the interaction between TMs and bottom graphene layer must yield the asymmetry of top and bottom layers. Actually, such kind of symmetry-lowering is present in the Co and Ni intercalated systems as we have mentioned above and the gap opening does occur. Of course, the gap size is sensitive to the distribution of intercalated TM atoms. Another key concern is the intercalated metal coverage. To illustrate this, we employ a $2\sqrt3 \times 2\sqrt3R30^{\circ}$ supercell but with one or two metal atoms removed to mimic the cases corresponding to different metal coverage. The results show that the gap size, either $E_g^u$ and $E_g^d$ for the Mn system or $E_g$ for the Li system, varies within $\pm$ 30 meV.

To summarize, we introduce a new paradigm to directly engineer an energy gap in bilayer graphene utilizing the local interaction or built-in electric field on semiconducting substrate. First-principles calculations indicate the gap size around 0.12-0.25 eV in non-covalently intercalated SiC systems, dependent upon the employed metals. A superior advantage of this scheme is the extremely high equivalent electric field, $\sim$4 V/nm, which enables the gap insensitive to the applied gate, behaving like an intrinsic one. We believe the fundamental principle underlying the use of metal decoration to realize the symmetry-breaking of bilayer graphene is generally applicable to other semiconducting substrates having appropriate surface configuration. These insights point to alternative and perhaps more realizable directions to incorporate graphene with current semiconductor industry for electronic applications.

We acknowledge the support of the National Natural Science Foundation of China (Grant Nos. 11304053).

\end{document}